# Evaluating the role of correlation among markers in prediction models

Sergio Sabroso-Lasa[1,*], Luis Mariano Esteban[2,3,*], Tomás Alcalá-Nalvaiz[4], Francisco J. Jurado[1], Núria Malats[1]

1 Genetic and Molecular Epidemiology Group, Spanish National Cancer Research Centre (CNIO) and CIBERONC, Madrid, Spain

2 Department of Applied Mathematics, Escuela Universitaria Politécnica de La Almunia, University of Zaragoza, Zaragoza, Spain

3 Institute for Biocomputation and Physics of Complex Systems (BIFI)

4 Department of Statistical Methods, University of Zaragoza, Zaragoza, Spain

* Corresponding authors:

-1) Sergio Sabroso-Lasa, Genetic and Molecular Epidemiology Group, Spanish National Cancer Research Centre (CNIO), C/ Melchor Fernández Almagro 3, 28029, Madrid, Spain; ssabroso@cnio.es; 2) Luis Mariano Esteban, Department of Applied Mathematics, Escuela Universitaria Politécnica de La Almunia, Universidad de Zaragoza, C/ Mayor 5, 50100, La Almunia de Doña Godina, Spain & Institute for Biocomputation and Physics of Complex Systems (BIFI), 50009, Zaragoza, Spain; lmeste@unizar.es

## Abstract:

**Backgorund**: different methods have been employed to estimate models to maximize their performance, i.e., the area under the receiver operating characteristic curve (ROC-AUC). Still, once a model incorporating different variables is developed, integrating novel biomarkers may improve the model's diagnostic ability. Unfortunately, the improvement in discrimination provided by resulting from adding a new biomarker into an existing model is not always evident, even if the new marker by itself has a good discrimination ability. The presence and sign of correlations between the biomarkers included in the model may impact the model's performance. In this paper, we assess the effect of the sign and magnitude of the correlations between markers in improving the discrimination ability of predictive models.

**Methods:** under multivariate normality assumption we have derived an expression for the maximum AUC as a function of the correlation between markers. Using surfaces, we show graphically the relationship between the increase of the AUC and the correlations. Logarithmic folded bivariate normal Gamma simulations have been performed in skewed data cases. Furthermore, the AUC improvement was assessed combining 1934 distinct blood lipid metabolites determined by liquid chromatography in 44 pancreatic cancer cases and 38 controls from the PanGenMic Study, conducted as real-data example.

**Results:** negative correlations consistently maximize the combined AUC in predictive models, offering significant improvements when markers have equal predictive ability (e.g., AUCs of 0.85, 0.70, and 0.60). In contrast, positive correlations yield the least favorable results. Negative correlations remain optimal for markers with differing abilities (e.g., AUC pairs of 0.90-0.70 and 0.70-0.55), though positive correlations show slight benefits. Simulations with skewed distributions confirm these trends, emphasizing the role of asymmetry and skewness (coefficients 1.5–2.5) in marker selection. Real-world analysis of serum lipid-derived metabolites for detecting pancreatic ductal adenocarcinoma (PDAC) reinforces the influence of correlations and distributions on AUC optimization.

**Conclusions:** the magnitude and sign of the correlations between biomarkers should be considered when selecting and including new markers in algorithms.

## 1. Background

The improvement in risk stratification, diagnosis, prognosis, and predictive risk algorithms in medicine is conditioned by the finding of the ideal biomarker, which is a variable that can optimally predict the occurrence of an event related to a disease. Medicine, at present, benefits of a variety of biomarkers with proven utility in different disease scenarios, but unfortunately, their prediction has room for improvement. Combining biomarkers to improve diagnostic accuracy is an issue of interest that does not have a global answer (1). The Receiver Operating Characteristic curve (ROC) and the area under the curve (AUC) have been used to measure the discriminatory ability of a predictive model(2). Although some markers have a reasonably good ability to discriminate between events and non-events, the combination of markers usually improves the overall diagnostic performance of the test (3,4).

One key point in this field is the effect of adding new markers to existing models. With numerical simulations, Pepe *et al.* demonstrated that statistical significance alone does not define a marker's discriminatory ability and that a meaningful AUC necessitates an association with a magnitude seldom observed in epidemiological studies (5). Substantive independent associations between the new marker and the outcome are necessary for models with standard risk factors and good discrimination to achieve a significantly larger AUC.

When a new biomarker is evaluated, in most cases, this new predictor variable is added to a set of standard features, so a comparison of models with or without the new variable added to the standard predictors is required; most times, the comparison is based on AUC bootstrap tests(6). What happens is that many of the newly suggested markers show no statistically significant difference between models using the AUC comparison. To make the comparisons more reliable, Pencina *et al*. proposed two indices (7), the net reclassification improvement (NRI) and the integrated discrimination improvement (IDI), to supplement the improvement in AUC, which may be too stringent to achieve. In an application for evaluating the incremental value of HDL cholesterol in heart disease risk models, they demonstrated that both NRI and IDI were highly significant in suggesting a significant improvement in performance, while the change in AUC disagreed. Unfortunately, Hilden *et al.(8)* showed that the NRI statistic can be significantly biased due to the use of miscalibrated risk models. Alternatively, Pinsky *et al.* have analyzed the role of correlations among markers in the discriminatory ability of predictive models (9). They conclude that adding markers significantly enhances the predictive ability of models, especially when they are negatively affected correlated. Also, Bansal *et al.* (10) used the optimal combination of markers under the bivariate binormal assumption to analyze how the correlation between markers leads to diagnostic improvement. In parallel with Pinsky *et al.* discoveries they found significantly larger gains in performance by adding a novel marker that performs poorly on its own but is highly correlated with the standard at least within one class category. Kundu *et al.* (11), through simulations that assume normality, also investigated the impact of correlations on both the training and external validation populations, finding a significant influence of correlations on the AUC of both populations depending on the sample size.

As a result of these studies, whether the sole criterion for selecting new markers in predictive models should be their individual performance is debatable. In addition, these studies assume normality and do not examine the implications of combining markers or scores exhibiting asymmetric distributions, which is very common in real-world data.

Furthermore, with the current advancements in technology, the vast amount of data generated, and the increase in computational capacity, a growing amount of data is being generated, and combining it appropriately can lead to much more accurate results. For instance, in oncology, obtaining the optimal combination of scores generated from various omics data (i.e., DNA, RNA, methylation, radiomics, imaging, etc.) along with non-omics data is becoming increasingly popular and crucial for improving the prediction of diagnosis or prognosis across different cancer types (12–15). Therefore, several clinical objectives could benefit if clear selection criteria, based on the study of the relationship between the markers to be combined and aimed at maximizing predictive capacity, are taken into account.

## 2. Methods

This study aims to analyze how the sign and magnitude of correlations between markers affect the discriminatory power of predictive models across scenarios that represent the diverse range of predictive markers observed in real-world data.

Assuming multivariate normality, we derive an expression for the maximum Area Under the Curve (AUC) as a function of the correlation between markers based on the results provided by Su and Liu(16). We analyzed the relationship between AUC improvement and the correlation's sign and magnitude using graphical representations. Simulations using logarithmic fold bivariate normal and gamma distributions were conducted to extend these findings to non-normal distributions. Lastly, the improvement in AUC was evaluated in real data by combining 1,934 distinct blood lipid metabolites, determined via liquid chromatography, in a practical example involving 44 pancreatic cancer cases and 38 controls from the PanGenEU Study.

*2.1 Prediction models under bivariate normal assumption*

Under the assumption of multivariate normality in both cases and controls, Su and Liu(16) expressed the maximum AUC for the linear combination of markers. When the multivariate normal distribution of diseased and non-diseased population is assumed, $N(\mu_1, \Sigma_1)$, $N(\mu_0, \Sigma_0)$, the area under the ROC curve of the optimal linear combination is given by the following expression:

$$AUC_{max} = \Phi\left(\sqrt{\mu^T(\Sigma_1 + \Sigma_0)^{-1}\mu}\right)$$

where $\Phi$ denotes the cumulative distribution function of the normal distribution, and the coefficients for the best linear combination are:

$$\beta_{max} \propto (\Sigma_1 + \Sigma_0)^{-1}\mu$$

where $\mu = \mu_1 - \mu_0$

Thus, if we have two markers, X the standard one and Y the novel marker, it is possible to calculate the maximum AUC corresponding to the best combination of X and Y, $AUC_{X,Y}$. Then, the improvement in discriminatory ability can be measured by the increase in AUC: $AUC_{X,Y} – AUC_X$

Bansal *et al.(10)* analyzed different scenarios for the binormal distribution (X, Y) with different correlations between X and Y for cases and controls. Here, we also analyze the binormal bivariate case. However, we consider a different approach, focusing the study on the relation between $AUC_{X,Y}$, and the sign and the magnitude of the correlation between markers. Henceforth, it can be assumed $\mu_0 = (0,0)$ for controls and $\mu_1 = (\mu_X, \mu_Y)$ for cases, and we denote the covariance matrices for cases and controls by $\Sigma_0, \Sigma_1$. The simplest case corresponds to independence $\rho_0 = \rho_1 = 0$ between X and Y in cases and controls. For simplicity, we took the marginal standard deviations as 1, then the covariance matrices are $\Sigma_0 = \Sigma_1 = \begin{pmatrix} 1 & 0 \\ 0 & 1 \end{pmatrix}$. Thus:

$$AUC_X = \Phi\left(\sqrt{\mu_X^2/2}\right), \qquad AUC_{X,Y} = \Phi\left(\sqrt{\frac{\mu_X^2 + \mu_Y^2}{2}}\right)$$

The formula shows that $AUC_{X,Y} - AUC_X$ is an increasing function of $\mu_Y$, therefore, an increasing function of $AUC_Y$. For this case, the discrimination ability of the novel marker Y is what causes the improvement in diagnostic accuracy.

In a more general scenario, assuming again $\mu_0 = (0{,}0)$, $\mu_1 = (\mu_X, \mu_Y)$ and choosing the correlation between X and Y in cases and controls as $\Sigma_0 = \begin{pmatrix} 1 & \rho_0 \\ \rho_0 & 1 \end{pmatrix}, \Sigma_1 = \begin{pmatrix} 1 & \rho_1 \\ \rho_1 & 1 \end{pmatrix}$, by operating with matrices, we can easily conclude that the maximum $AUC_{X,Y}$ is:

$$AUC_{X,Y} = \Phi\left(\sqrt{\frac{2\mu_X\mu_Y(\rho_0 + \rho_1) - 2\mu_X^2 - 2\mu_Y^2}{(\rho_0 + \rho_1 + 2)(\rho_0 + \rho_1 - 2)}}\right)$$

This is a function that depends on four variables $\mu_X$, $\mu_Y$, $\rho_0$, $\rho_1$.

From this point, straightforward expressions can be derived for specific scenarios:

For example, when $\mu_X = \mu_Y = \mu$ this formula is reduced to:

$$AUC_{X,Y} = \Phi\left(\sqrt{\frac{2\mu^2(\rho_0 + \rho_1) - 4\mu^2}{(\rho_0 + \rho_1 + 2)(\rho_0 + \rho_1 - 2)}}\right) = \Phi\left(\sqrt{\frac{2\mu^2(\rho_0 + \rho_1 - 2)}{(\rho_0 + \rho_1 + 2)(\rho_0 + \rho_1 - 2)}}\right) = \Phi\left(\sqrt{\frac{2\mu^2}{\rho_0 + \rho_1 + 2}}\right) \quad (1)$$

and in cases when $\mu_X \neq \mu_Y$, but $\rho_0 = \rho_1 = \rho$, then:

$$AUC_{X,Y} = \Phi\left(\sqrt{\frac{4\mu_X\mu_Y\rho - 2\mu_X^2 - 2\mu_Y^2}{(2\rho+2)(2\rho-2)}}\right) = \Phi\left(\sqrt{\frac{2\mu_X\mu_Y\rho - \mu_X^2 - \mu_Y^2}{2\rho^2-2}}\right) = \Phi\left(\sqrt{\frac{\mu_X^2 + \mu_Y^2 - 2\mu_X\mu_Y\rho}{2 - 2\rho^2}}\right) \quad (2)$$

For the remaining cases with different values of $\mu_X$ and $\mu_Y$ corresponding to weak or strong markers, we were able to analyze through simulations the relation between $AUC_{X,Y,}$ and $(\rho_0, \rho_1)$ by means of surfaces where the X and Y axis correspond to $(\rho_0, \rho_1)$ and the Z axis to $AUC_{X,Y}$.

First, we considered and simulated X and Y as markers, variables with the same predictive capacity, with AUC values of 0.6, 0.7, and 0.85. The correlation in cases and controls is displayed on the X and Y axes, both from -0.99 to 0.99 by 0.01, while the Z axis shows the AUC of the marker created by the optimal linear combination of the X and Y markers. Second, we examine markers with different discriminatory capacities, specifically the pairs (0.90, 0.50), (0.90, 0.70), and (0.70, 0.55).

*2.2 Prediction models under skewness distributions*

While the role of correlation in the combination of markers or omics scores is an issue that has not yet been solved globally, all studies that analyze this topic are based on scenarios where normality is assumed(17,18). Markers associated with prediction in oncology are often not symmetric, as they tend to show extreme values in cases of disease progression or advanced staging. In addition to normality, this paper analyzes simulation scenarios where the variables to be combined follow asymmetric distributions, particularly in the case of the population.

For this purpose, two different asymmetric distributions have been considered: logarithmic folded bivariate normal variables, with a not very high skewness, and gamma variables, with a much more marked skewed component.

### 2.2.1 Logarithmic folded bivariate normal distributions

As previously mentioned, various asymmetric scenarios have been simulated. Firstly, with the aim of obtaining a mildly pronounced asymmetric component, we assume that the combined variables X and Y follow a logarithmic folded bivariate normal distribution in both cases and controls. Specifically, with the notation described earlier, this means:

$$\begin{pmatrix} X_0 \\ Y_0 \end{pmatrix} \equiv log\left( abs\left( N\left( \begin{pmatrix} \mu_{X_0} \\ \mu_{Y_0} \end{pmatrix}, \begin{pmatrix} 1 & \rho_0 \\ \rho_0 & 1 \end{pmatrix} \right) \right) \right) for\ controls,$$

$$\begin{pmatrix} X_1 \\ Y_1 \end{pmatrix} \equiv log\left( abs\left( N\left( \begin{pmatrix} \mu_{X_0} \\ \mu_{Y_0} \end{pmatrix}, \begin{pmatrix} 1 & \rho_1 \\ \rho_1 & 1 \end{pmatrix} \right) \right) \right) for\ cases.$$

Since, in this case, there is no analytical formula to provide the maximum obtained AUC, a stepwise variable combination algorithm (19) has been used to obtain the AUC of the combined model. Fisher's skewness coefficient was analytically calculated for each variable and scenario to measure the non-normality of the different simulations. Note that the absolute value of the simulated values was considered to avoid taking the logarithm of negative values. Additionally, a small constant was added to all simulated values to ensure they are always greater than zero.

Regarding the simulations, the *mvrnorm()* function from the *MASS* package was used. This function allowed the generation of values from a bivariate normal distribution when provided with the mean vectors $\mu_0$, $\mu_1$, and the variance-covariance matrices $\Sigma_{0,}$ $\Sigma_1$ as parameters. Therefore, by fixing the mean vectors, we could simulate the variables X and Y for both cases and controls, exploring all possible combinations of correlations between them. This enabled us to assess how the increase in $AUC_{X,Y}$ that vary according to these correlations.

For each possible combination, bivariate normal distributions are generated for 100 cases and 100 controls ($n = 200$) to avoid imbalance problems(20). The results were obtained by calculating the mean across 1000 repetitions to mitigate potential biases due to randomness.

Similarly to the normal distributions, markers with the same predictive ability were examined, ($AUC_X = AUC_Y = 0.85$; $AUC_X = AUC_Y = 0.70$), and combinations of different predictive capacities (0.85, 0.70), (0.85, 0.55). Since no analytical formula is available as in the previous case, the graphical results are presented using heatmaps.

### *2.2.2 Gamma distributions*

Different simulations have been considered to obtain distributions with a more pronounced asymmetric component compared to the previous case where the variables X and Y follow a gamma distribution(21,22) for cases and a normal distribution for controls. Specifically, the following scenario is proposed:

$$\begin{pmatrix} X_0 \\ Y_0 \end{pmatrix} \equiv N\left(\begin{pmatrix} \mu_{X_0} \\ \mu_{Y_0} \end{pmatrix}, \begin{pmatrix} 1 & \rho_0 \\ \rho_0 & 1 \end{pmatrix}\right) \text{ } for\ controls,$$

$$\begin{pmatrix} X_1 \\ Y_1 \end{pmatrix} \equiv \begin{pmatrix} \Gamma\left(\alpha_{X_1}, \lambda_{X_1}\right) \\ \beta_1 X_1 + \beta_2 \Gamma\left(1, \lambda_{Y_1}\right) \end{pmatrix} \text{ } for\ cases.$$

where $\alpha_{X_1}, \lambda_{X_1}, \lambda_{Y_1}$ are the shape and scale parameters of the gamma distributions, respectively, and $\beta_1, \beta_2 > 0$ scalar parameters to be defined across different scenarios. Note that controls follow a normal distribution to manage correlations to be estimated between the generated variables. Additionally, the variable $Y_1$ is a linear combination of $X_1$ to preserve both the desired correlations and the asymmetric component.

The simulation parameters are the same as those used in the logarithmic folded bivariate normal case. Since an analytical formula to determine the maximum $AUC_{X,Y}$ was not available, a step-by-step algorithm, which is employed once again to identify the optimal linear combination of the markers. The results are presented using heatmaps.

*2.3 Real world-data practical example: Metabolome in pancreatic cancer patients*

In addition to the analytical analysis and simulations, a real data set was considered to evaluate if the hypotheses obtained in theory were extensible to practice. For this purpose, the AUC improvement was assessed by combining 1,934 distinct blood lipid metabolites determined by liquid chromatography in 44 pancreatic ductal adenocarcinoma (PDAC) cases and 38 controls recruited from two hospitals in Spain (23) between 2016 and 2018 participating in the PanGenEU Study were considered here. Eligible cases were patients aged >18 years with a confirmed diagnosis of PDAC. Controls were aged- and sex-matched (±10 years) patients from the same hospital catchment area but hospitalized for reasons unrelated to PDAC (24). All subjects were informed about the study´s aims and protocols, and they obtained written consent. Independent ethics committees of the participating hospitals approved the studies. The characteristics of the PanGenEU population were described as follows: Continuous variables were described using the median and interquartile range, while categorical variables were expressed as absolute and relative frequencies. Differences by case/control were assessed using the Mann-Whitney or chi-squared tests, as appropriate, and p-values $\leq 0.05$ were considered statistically significant.

In this practical case, metabolites with varying individual predictive capacities were selected and combined with each of the remaining available metabolites using two-variable logistic regression models. The goal is to examine how the combined model's AUC improves depending on the second marker's individual predictive capacity and the correlation between them.

### *2.4 Software specifications*

Simulations, statistical analyses, and visualizations for all datasets were conducted using R Statistical Software v.4.4.2 (The R Foundation for Statistical Computing, Vienna, Austria).

## 3. Results

This section outlines the results obtained graphically in both normality and asymmetry scenarios and in the real-world example of detecting PDAC patients using metabolites derived from serum lipids.

### *3.1 Bivariate normal distributions*

Under the assumption of bivariate normality in cases and controls and based on the formula proposed by Su and Liu(16), we have derived expressions analytically, demonstrating how the maximum AUC depends on the sign and magnitude of the correlations between cases and controls. We observed that negative correlations significantly increase the combined AUC. The surfaces generated to represent the generic case visually are shown in **Figure 1**.

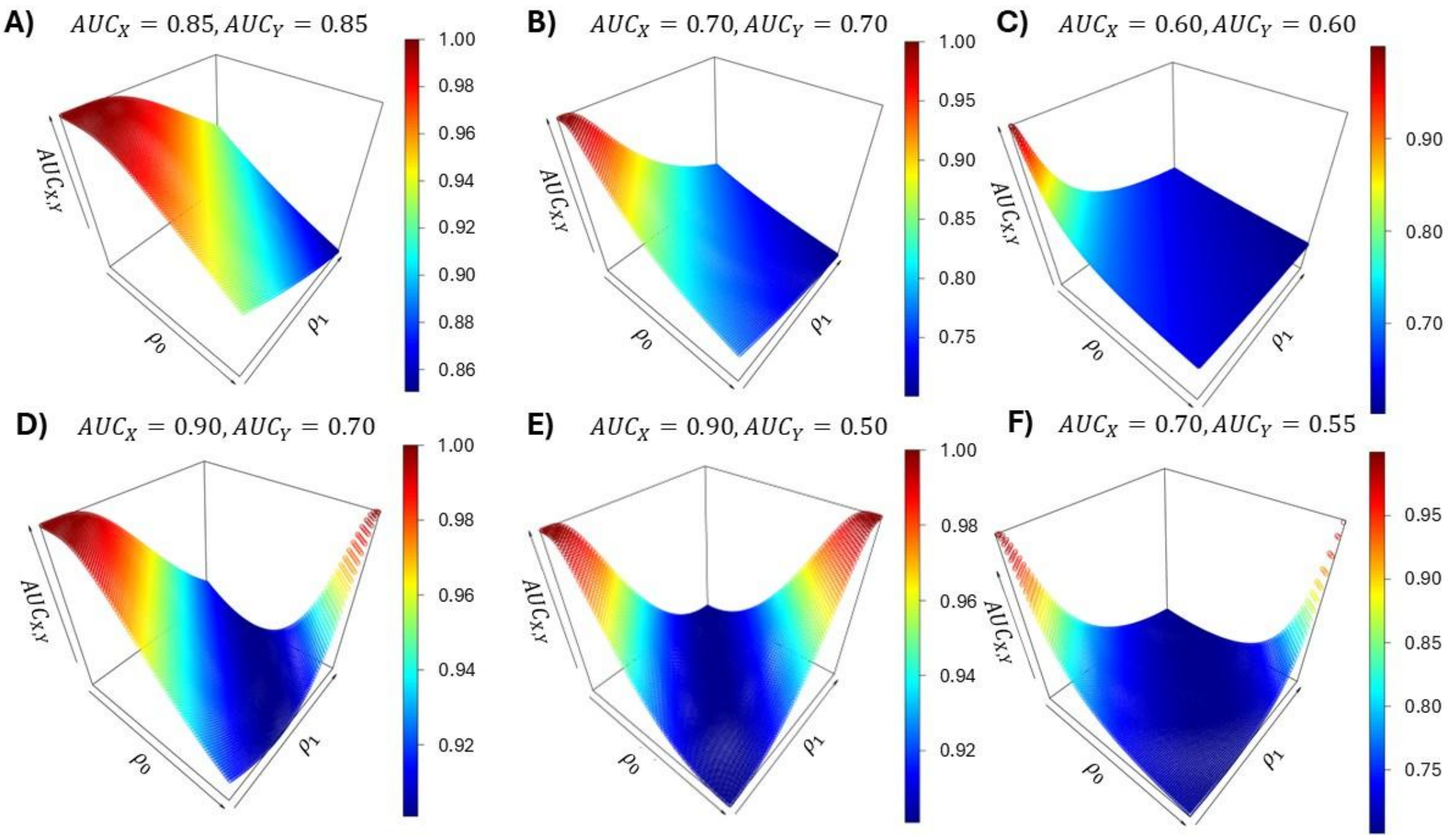

**Figure 1**: Simulation results for the bivariate normal case presented as surfaces. The x and y axes represent the correlations of the markers to be combined in cases and controls. In contrast, the z-axis represents the maximum combined AUC obtained using the Su and Liu formula. The surfaces in the top row (1A, 1B, 1C) depict scenarios where the markers to be combined have the same predictive ability. In contrast, the three surfaces in the bottom row (1D, 1E, 1F) illustrate scenarios where markers with differing individual AUCs are combined.

On one hand, for scenarios **1A**, **1B**, and **1C**, where both markers X and Y to be combined have the same predictive ability (0.85, 0.70, and 0.60, respectively), it is evident that the combined AUC reaches its maximum when X and Y are negatively correlated in both the cases and controls. Furthermore, the combined AUC rises as both separated magnitudes increase. Moreover, when the markers are independent, a notable improvement in the combined AUC is still evident, with positive correlations indicating the worst-case scenario scenario. On the other hand, in scenarios **1D**, **1E**, and **1F**, where the markers to be combined have different predictive abilities—specifically the pairs (0.90, 0.70), (0.90, 0.50), and (0.70, 0.55), respectively—negative correlations remain the ones yielding the maximum combined AUC. However, positive correlations play a more significant role, and independence between markers becomes the least favorable in these cases.

*3.2 Skewed distributions case. Simulations*

Multiple log-normal and gamma distribution simulations were conducted for scenarios where normality was not assumed.

**Figure 2** shows the heatmaps generated assuming that both cases and controls follow log-normal distributions. As in the normality case, scenarios were considered where the markers X and Y have the same predictive ability (**2A**, **2B**, with AUCs of 0.85 and 0.71, respectively), as well as combinations of different individual AUCs (**2C** and **2D**, with the pairs 0.85-0.71 and 0.85-0.55, respectively).

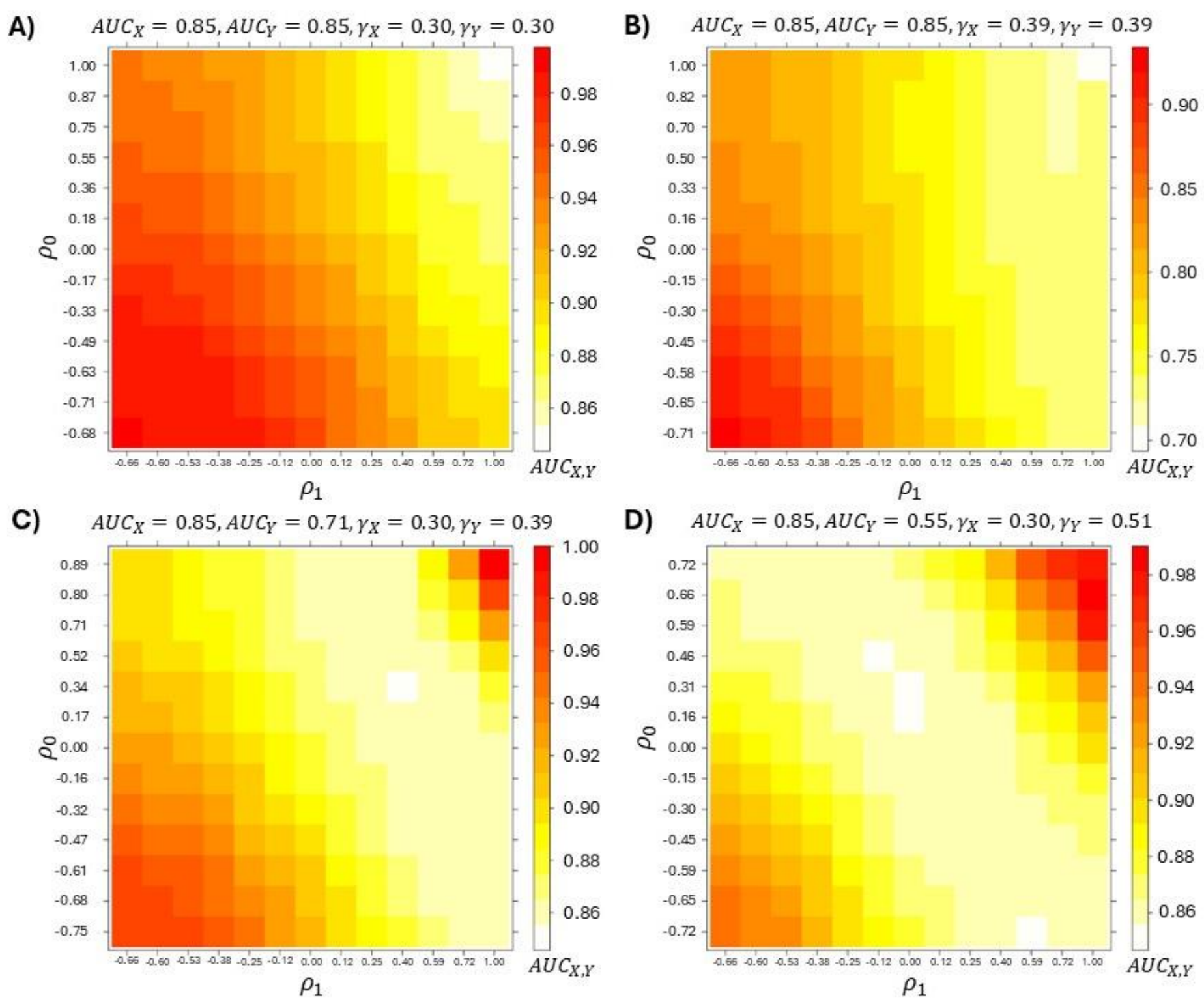

**Figure 2**: Heatmaps resulting from the simulation for the logarithmic folded bivariate normal scenario in cases and controls presented as heatmaps. The x-axis represents the correlation in cases, while the y-axis represents the correlation in controls. Color intensity indicates the increase in predictive performance of the combined model for these markers using the step-by-step prediction algorithm proposed by Esteban et al. The top heatmaps (2A, 2B) depict scenarios where the markers to be combined possess the same predictive ability, while the bottom heatmaps (2C, 2D) represent scenarios with differing predictive abilities. The analytical Fisher skewness coefficient of both simulated markers is indicated in the title of each heatmap.

Similarly to the surfaces, it can be observed that in cases of equal predictive ability (A, B), the maximum combined AUC is achieved when the correlations are negative, with situations closer to -1 in both cases and controls yielding the highest

AUC. The improvement gradient decreases as we approach marker independence, reaching the lowest values when both cases and controls exhibit positive correlations. However, this pattern changes when the individual predictive abilities differ, as in the normality case. Some improvement is observed when the markers have very positive correlations, while correlations close to 0 do not enhance the individual models. In any case, negative correlations provide the best improvements. Nonetheless, as seen in the four scenarios analyzed, the Fisher skewness coefficients obtained for our markers X and Y, which follow log-normal distributions, are relatively low (ranging from 0.3 to 0.5). Therefore, more pronounced asymmetrical scenarios are still required for more significant effects.

To address this, scenarios were examined in which gamma distributions are assumed for cases and normal distributions for controls. The variation in the combined AUC under these conditions, as correlations change, is shown in **Figure 3**.

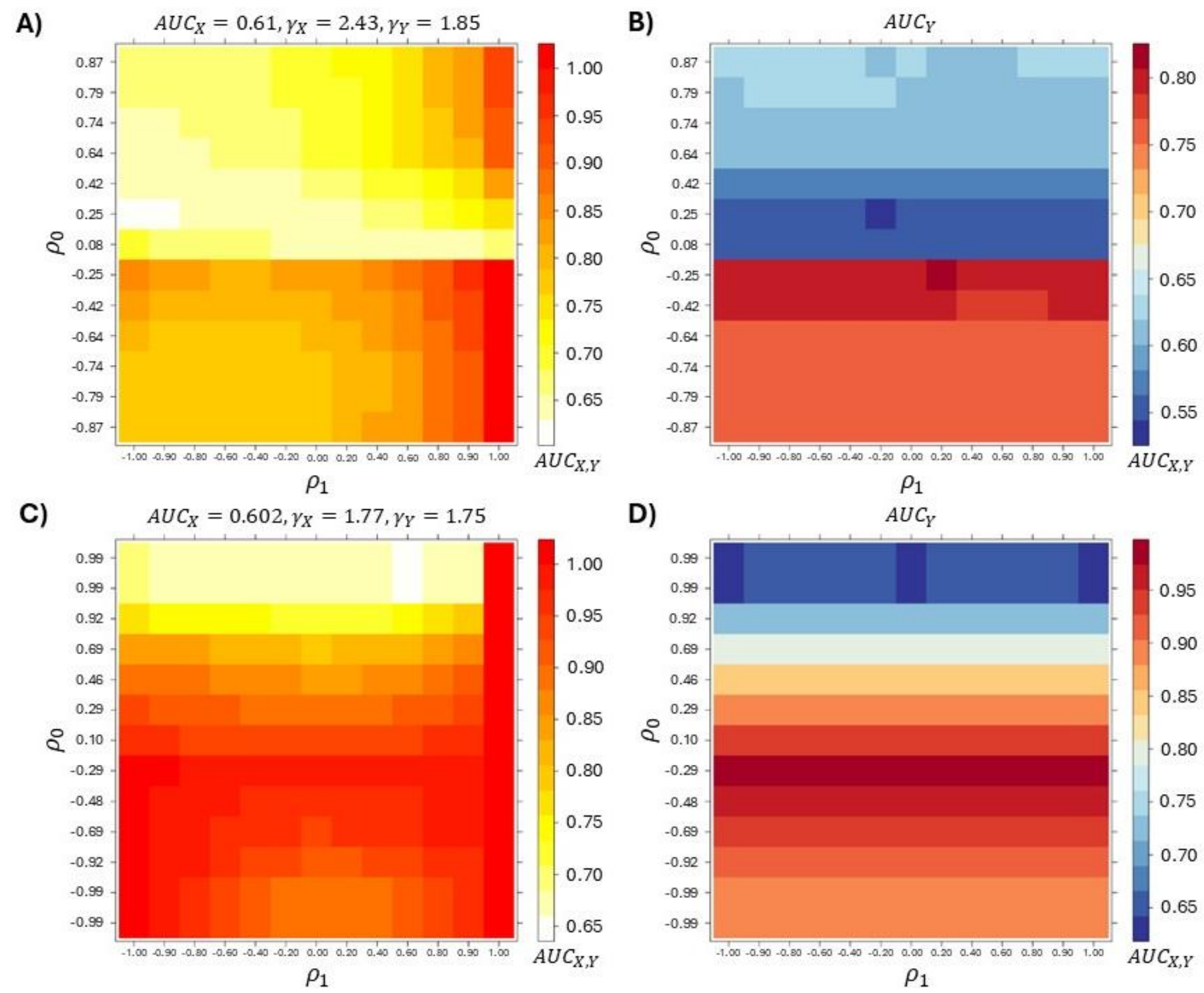

**Figure 3**: Simulation results for a gamma distribution in cases and a normal distribution in controls are presented as heatmaps. The x-axis shows the correlation in cases, while the y-axis shows the correlation in controls. The heatmaps on the left illustrate the increase in the combined AUC, whereas, due to the complexity of fixing AUCs and achieving all possible correlations in these scenarios, the heatmaps on the right display how the individual AUC of the second marker varies. Color intensity reflects the improvement in the predictive performance of the combined model for these markers, based on the step-by-step prediction algorithm proposed by Esteban et al. The top row of heatmaps (3A) corresponds to scenarios where the markers to be combined have identical predictive abilities, while the bottom row (3B) represents scenarios with differing predictive abilities. The simulated markers' Fisher skewness coefficient is indicated in each heatmap's title.

Under these distributions, it is evident that the skewness coefficients vary in range (1.5, 2.5), which is significantly higher than when logarithmic folded bivariate normality is assumed. However, due to the distributions of markers X and Y, it is not possible to independently control the individual AUC of the second marker and the correlations between them, so we also included the $AUC_Y$ each case on the left heatmaps. As in other cases, markers with similar predictive ability (**3A**, $AUC_X = 0.608$, $AUC_Y = 0.674$) and different predictive abilities (**3B**, $AUC_X = 0.602$, $AUC_Y = 0.851$) were considered. However, the results are not as easy to interpret as before, and no clear conclusions can be obtained regarding which correlation sign favors the combined AUC. This suggests that asymmetry between the variables to be combined is a crucial factor when selecting markers.

### *3.3 Real world-data practical example*

Clinico-demographical characteristics of the PanGenEU population by case-control are shown in **Table 1**.

| | All N=116 | Cancer N=60 | Control N=56 | p-value |
|---|---|---|---|---|
| Age | 72.0 (11.3) | 71.9 (10.4) | 72.0 (12.2) | 0.976 |
| Gender: | | | | 0.759 |
| Female | 47 (40.5%) | 23 (38.3%) | 24 (42.9%) | |
| Male | 69 (59.5%) | 37 (61.7%) | 32 (57.1%) | |
| Centre: | | | | 0.775 |
| Hospital 1 | 73 (62.9%) | 39 (65.0%) | 34 (60.7%) | |
| Hospital 2 | 43 (37.1%) | 21 (35.0%) | 22 (39.3%) | |
| Smoking Status: | | | | 0.744 |
| Current | 16 (14.0%) | 9 (15.5%) | 7 (12.5%) | |
| Never | 57 (50.0%) | 30 (51.7%) | 27 (48.2%) | |
| Occasional | 41 (36.0%) | 19 (32.8%) | 22 (39.3%) | |
| Alcohol Status: | | | | 0.199 |
| Current | 83 (72.2%) | 39 (66.1%) | 44 (78.6%) | |
| Never | 32 (27.8%) | 20 (33.9%) | 12 (21.4%) | |
| Diabetes: | | | | 0.007 |
| ≥ 2 years | 17 (14.9%) | 11 (18.6%) | 6 (10.9%) | |
| < 2 years | 7 (6.14%) | 7 (11.9%) | 0 (0.00%) | |
| Not diagnosed | 90 (78.9%) | 41 (69.5%) | 49 (89.1%) | |
| Obesity: | | | | 0.940 |
| Not Obese | 81 (73.0%) | 43 (74.1%) | 38 (71.7%) | |
| Obese | 30 (27.0%) | 15 (25.9%) | 15 (28.3%) | |

**Table 1**: Clinicodemographic characteristics of the PanGenEU subset population. Stratified analysis by case-control status. P-values obtained from Mann-Whitney-Wilcoxon tests for continuous variables or Chi-square tests for categorical variables, as appropriate.

The descriptive table indicates that there are no significant differences between cases and controls in the clinical-demographic variables age, gender, centre, smoking and alcohol status, and obesity. However, it shows significant differences in diabetes mellitus (p = 0.007), with all cases relating to patients diagnosed with new-onset diabetes mellitus (diagnosed within the last 2 years).

In this practical assessment, metabolites with different individual predictive abilities were selected, and the variation in the combined AUC of the step-by-step model proposed by Esteban *et al.*(19) with two metabolites analyzed based on the individual predictive ability of the second marker and the correlation between both markers. **Figure 4** shows the results obtained from 4 metabolites with individual AUCs of 0.808, 0.708, 0.603, and 0.5, respectively, with the first metabolite (**4A**) having the highest individual AUC among the 1,934 available.

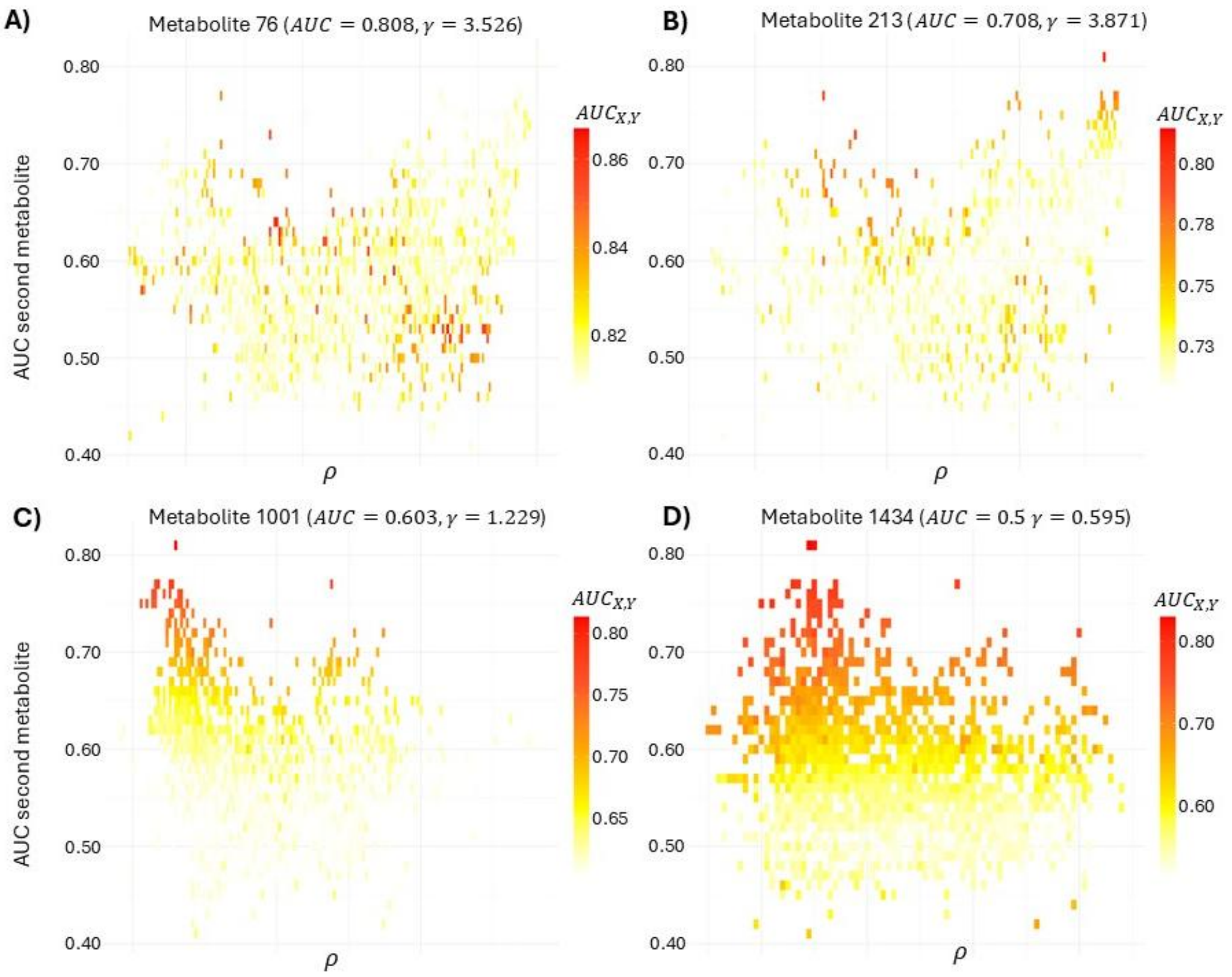

**Figure 4**: Results from the real-world practical case predicting the presence or absence of PDAC using lipid metabolites extracted through liquid chromatography. Metabolites with varying individual predictive abilities were selected, and pairwise combinations with other available metabolites were analyzed using the step-by-step algorithm proposed by Esteban et al. The x-axis represents the correlation between the two selected metabolites, while the y-axis shows the predictive ability of the second metabolite. Color intensity reflects the increase in the combined AUC. Metabolites with different Fisher skewness coefficients were used for this analysis: the top panels (4A, 4B) display metabolites with a strongly pronounced skewness, while the bottom

panels (4C, 4D) represent metabolites with a less pronounced but still high skewness component.

In general, for case **4A** (metabolite 76, the one with the highest predictive ability), the upper part of the plot shows that markers with greater individual predictive abilities, and thus more similar AUCs to the initial metabolite, result in a greater increase in the combined AUC when they are negatively correlated. However, in the lower section of the plot, where markers exhibit AUCs different from the initial metabolite, positive correlations start to gain significance, resulting in the highest overall AUC increases. In addition to these patterns, the plot also features regions with high combined AUCs that make its interpretation more complex. This complexity is further influenced by the highly asymmetric nature of the original marker, since $\gamma_X = 3.526$. A similar situation is observed in case **4B**, where, despite some general trends, the results remain abstract, likely due to the high asymmetry of this metabolite ($\gamma_X = 3.871$). In contrast, for metabolites with lower asymmetry (cases **4C** and **4D**, with $\gamma_X = 1.229, 0.595$ , respectively), a much clearer and more structured trend emerges. In both scenarios, there is a noticeable gradient of improvement favoring negative correlations, aligning with the results from our analytical analyses, surface plots, and simulations

## 4. Discussion

In this study, we analyzed the influence of correlations between cases and controls for combining two markers on the discriminatory ability of prediction models. Firstly, and as in most published studies(6,7,10,16,25), we assessed the case assuming bivariate normality. However, to the best of our knowledge, we have demonstrated this analytically for certain specific cases for the first time in the literature. Specifically, through a series of operations and based on the formula by Su and Liu(16), we observed that when $\mu_X$ = $\mu_Y$ = $\mu$:

$$AUC_{X,Y} = \Phi\left(\sqrt{\frac{2\mu^2}{\rho_0 + \rho_1 + 2}}\right) \quad \textbf{(Eq. 1)}$$

and in cases when $\mu_X \neq \mu_Y$, but $\rho_0 = \rho_1 = \rho$, then:

$$AUC_{X,Y} = \Phi\left(\sqrt{\frac{\mu_X^2 + \mu_Y^2 - 2\mu_X\mu_Y\rho}{2 - 2\rho^2}}\right) \quad \textbf{(Eq. 2)}$$

Note that, since $\Phi$ and $\sqrt{a}$ are both monotonically increasing functions, $\Phi(\sqrt{a})$ increases as $a$ increases.

In both simplified cases, we can analytically observe that the $AUC_{X,Y}$ will be maximized when the correlations are negative, since in case **(Eq. 1)**, $AUC_{X,Y}$ will reach its maximum when the denominator is as small as possible, i.e., when $\rho_0 + \rho_1 \approx -2$. This condition is only met when both $\rho_0, \rho_1$ are negative and in case **(Eq. 2)**, since $\rho$ in the denominator is squared, its sign does not affect the value of $AUC_{X,Y}$. Therefore, only the numerator will impact the model performance. In this situation, maximizing the numerator is optimal, which occurs when $\rho$ is negative, as it is multiplied by *-2*. Nevertheless, in this scenario, the mean vectors also play a significant role, which necessitates separate and detailed analysis.

For the general case, which we analyzed through the simulation of surfaces (**Figure 1**), we observed how negative correlations are the most relevant to improve the combined AUC, whereas the independence and positive correlations vary depend on the predictive ability of the markers to be combined. Independence is more relevant in scenarios with equal predictive ability, while positive correlations become more significant when combining markers with different AUCs. This can also be easily deduced from the expressions derived from the formula by Su and Liu(16): In the case of equal predictive ability (**Eq. 1**), and based on the same reasoning as before, we aimed to make the denominator as small as possible. Therefore, independence between markers in cases and controls is a more favorable situation than positive correlations. When the predictive abilities are different (**Eq. 2**), note that we have the squared correlation in the denominator, and therefore, its sign does not matter, only its magnitude. Notice that the fact that it is multiplied by -2 in the numerator guides negative correlations as the most favorable case. However, as mentioned earlier, positive correlations increase the combined AUC more than the independence factor, making it the worst-case scenario.

To the best of our knowledge, all published analyses reviewed to date assume normal distributions in their simulations. However, none of them address scenarios where the distributions of the markers to be combined in cases and controls are non-normal. In this study, we analyzed different non-normal distributions and evaluated, through simulations and a real practical case, whether the applicability of previous conclusions depends on the degree of skewness observed. The results were consistent with those previously reported for log-normal distributions or real metabolites with mild skewness. However, this interpretability is lost for gamma-distributed markers/scores or metabolites with significantly higher skewness coefficients, especially in cases.

In any case, it is worth noting that in all scenarios, combining the two markers with the highest individual AUCs does not always yield the greatest improvement. In some instances, as demonstrated by Pinsky & Zhu in a real example of ovarian cancer(9), adding a new feature with a low AUC can instead maximize the model's predictive ability. In this example, using CA125 as the primary marker, they observed that among the 28 available variables, the one achieving the greatest increase in the combined AUC was the only one negatively correlated with CA125. Notably, this variable had an individual AUC of 0.598, significantly lower than others in the dataset (AUC around 0.8).

However, a drawback to these findings is that identifying biomarkers that are negatively correlated in practice is not simple. For instance, in the previously mentioned ovarian cancer example by Pinsky and Zhu(9), among the top 14 biomarkers proposed to complement CA125, only 3 exhibited negative correlations in cases (21.43%) and 4 in controls (28.57%). Moreover, only one biomarker, IGF-II, showed negative correlations in both cases and controls simultaneously. Although with smaller differences, a similar trend is observed in our practical case. From the correlation matrix generated by pairwise comparisons of all possible correlations among the 1,934 serum lipid metabolites, we found 42.2% negative correlations in cases (788,764/1,869,211), 42.1% in controls (786,042/1,869,211), and 42.4% overall (793,431/1,869,211).

One of the main aspects to consider regarding these results is whether they can be extrapolated to combining more than two biomarkers. Investigating how correlations influence the increase in the AUC of a predictive model when considering trios of biomarkers becomes complex and laborious. A possible solution to this issue is to adopt a sequential approach: Once the first biomarker of the model is selected, a second potentially optimal biomarker is identified based on its correlation in cases and controls. Subsequently, these two biomarkers are combined into a single variable, and a third biomarker is sought that is optimal concerning its correlation with this new combined variable. Ultimately, this process can be repeated until the number of biomarkers is suitable for the available sample size, thereby preventing overfitting. This approach allows us to tackle the issue of two variables and even merge various scores generated by previously defined variable combinations.

Advances in technology and the increasing number of generated variables suggest that the current focus is less on the quantity of a number of variables and more on how they are combined. Increasing the number of variables is not always beneficial, and including all of them in a model does not necessarily lead to better outcomes(26). Following the principle of parsimony, where simpler models should always be preferred when performance is comparable, it is often more advantageous to include a smaller number of well-selected variables(27). Therefore, conducting a prior study of the variables to be included in a model is always valuable, as the "what" is combined seems to be more critical than the "how much" is combined.

This underscores the necessity for additional studies to establish criteria for optimal variable selection. In this manuscript, we show that examining the correlations of variables in cases and controls before their inclusion in a model can greatly improve its predictive ability while simplifying it compared to what is generally found in the current literature.

## 5. Conclusions

This manuscript demonstrates that correlations between features play a fundamental role in enhancing predictive models. Negative correlations should be taken into account in cases of normality. For skewness, a prior transformation is recommended before integration to support the earlier conclusion.

## 6. List of abbreviations

Area under the receiver operating characteristic curve (ROC-AUC); Pancreatic ductal adenocarcinoma (PDAC); Net reclassification improvement (NRI); Integrated discrimination improvement (IDI).

## 7. Declarations

### Ethics approval and consent to participate

IRB ethical approval (CEI PI 26 2015-v7) and written informed consent were obtained from participating centres and study participants, respectively. Clinical trial number: not applicable.

### Consent for publication

Not applicable.

### Availability of data and materials

Data is available upon reasonable request.

### Competing interests

The authors declare that the research was conducted without any commercial or financial relationships that could potentially create a conflict of interest.

### Funding

The study was partially funded by World Cancer Research [#15-0391], Fondo de Investigaciones Sanitarias (FIS), Instituto de Salud Carlos III-FEDER, Spain [Grant numbers PI15/01573, PI18/01347, FIS PI17/02303]; III beca Carmen Delgado/Miguel Pérez-Mateo de AESPANC-ACANPAN; and the EU-6FP Integrated Project [#018771-MOLDIAG-PACA]; EU-FP7-HEALTH [#259737-CANCERALIA]; Implementación en hospitales terciarios del algoritmo IA-PMPD para la predicción de metástasis de cáncer de páncreas y demostración de su rendimiento en tiempo real, funded by NextGenerationEU and Secretaría de Estado de Digitalización e Inteligencia Artificial del Ministerio para la Transición Digital y la Administración.  The corresponding author acknowledges funding from the I+D+I TALENTO DIGITAL NextGen program, supported by the European Union – NextGenerationEU [C005/24-ED CV1]. In addition, both corresponding authors S.S-L and L.M.E received funding from Proyectos de generación de conocimiento 2023 [PID2023-150234NB-I00] and Grupo T69_23R del Gobierno de Aragón.

### Author's contributions

S.S-L., L.M.E., N.M. and T.A. conceived and led the study. S.S-L. and L.M.E. did the analytical procedures and the corresponding simulations while N.M. and T.A. were crucial in their interpretation. N.M., S.S-L. and F.J.J. collected and curate the data. S.S-L., L.M.E., N.M. and T.A. wrote the manuscript and its corresponding reviews. L.M.E., N.M. and T.A. supervised the whole study.

### Acknowledgements

Not applicable.